\pdfoutput=1
\documentclass[english,prb,balancelastpage,groupedaddress,twocolumn,aps,10pt,floatfix]{revtex4-1}
\usepackage[T1]{fontenc}
\usepackage[latin9]{inputenc}
\usepackage{babel}
\usepackage[caption=false]{subfig}
\usepackage{amsmath}
\usepackage{amssymb}
\usepackage{graphicx}
\usepackage{dsfont}
\usepackage[unicode=true,pdfusetitle,
 bookmarks=true,bookmarksnumbered=false,bookmarksopen=false,
 breaklinks=false,pdfborder={0 0 1},backref=false,colorlinks=false]
 {hyperref}

\begin{document}

\title{Edge Majoranas on locally flat surfaces - the cone and the M\"obius band}
\author{A. Quelle}

\affiliation{Institute for Theoretical Physics, Center for Extreme Matter and
Emergent Phenomena, Utrecht University, Leuvenlaan 4, 3584 CE Utrecht,
The Netherlands}

\author{T. Kvorning}

\affiliation{Department of Physics, Stockholm University, AlbaNova University
Center, SE-106 91 Stockholm, Sweden}

\author{T.H. Hansson}

\affiliation{Department of Physics, Stockholm University, AlbaNova University
Center, SE-106 91 Stockholm, Sweden}

\author{C. \surname{Morais Smith}}

\affiliation{Institute for Theoretical Physics, Center for Extreme Matter and
Emergent Phenomena, Utrecht University, Leuvenlaan 4, 3584 CE Utrecht,
The Netherlands}

\date{\today}
\begin{abstract}
In this paper, we investigate the edge Majorana modes in the simplest
possible $p{}_{x}+ip_{y}$ superconductor defined  on surfaces with different geometry - 
the annulus, the cylinder, the M\"obius band and a cone (by cone we mean a cone with the tip cut away so it is topologically equivalent to the annulus and cylinder)- and with
different configuration of magnetic fluxes threading holes in these surfaces. 
In particular, we shall address two questions:
Given that, in the absence of any flux, the ground state on the annulus does not support Majorana modes, 
while the one on the cylinder does,
how is it possible that the conical geometry can interpolate smoothly between the two? 
Given that in finite geometries edge Majorana modes  have to come in pairs, how can a  $p{}_{x}+ip_{y}$ state be defined on a  M\"obius band, which has only  one edge? We show that the key to answering  these questions is that the ground state depends on
the geometry, even though all the surfaces are locally flat. In the case of the truncated cone, there is a non-trivial holonomy, while 
the  non-orientable M\"obius band must necessarily support a domain wall.

\end{abstract}
\maketitle

\section{Introduction}

Among all the fascinating features that topological states of matter exhibit, excitations with non-Abelian statistics are one of the most intriguing. They were first predicted to occur in quantum Hall systems at filling fraction $\nu=5/2$, $\nu=7/2$ (see Refs. \onlinecite{moore91,read96}), and later at other fractions in higher Landau levels\cite{read99}. In addition, they are expected to arise in 2d spinless $p_{x}+ip_{y}$ superconductors, where the vortices can be shown to obey non-Abelian Ising statistics \cite{read00,greiter92,ivanov01,stern04}. The anyonic nature of the vortices in such a superconductor originate from the presence of zero-energy Majorana modes in the vortex cores (in what follows, Majorana will always refer to a zero-energy Majorana mode). Such Majorana modes have not yet been observed in 2d systems, but there is experimental evidence pointing to their existence in 1d topological superconductors \cite{kouwenhoven12,marcus13}.

In 2d superconductors, the presence of such Majorana modes is determined by the amount of flux piercing the superconductor at a vortex: an odd number of superconducting flux quanta (unit flux) gives a Majorana, while an even number does not. Since Majorana modes can only come in pairs, an odd total number of unit flux vortices requires the presence of a Majorana mode at the boundary of the material. To understand this point, it is important to realize that a topological superconductor always has gapless edge modes in the thermodynamic limit, but there is not necessarily an edge mode precisely at zero energy (an edge Majorana) for finite edge length\cite{read00}. To know the number of Majorana modes in the presence of an arbitrary number of unit fluxes, one needs to know how many Majorana modes are present in the absence of flux. If the boundary consists of multiple edges, an even number of edge Majoranas might be present in the absence of flux.

To answer this question, one has to realise that the $p$-wave order parameter is not, as in the $s$-wave case, just a scalar field. Rather, it is a vector field, and, consequently, it couples to the geometry of the surface. At first, one would think that only the local curvature of the surface would matter, so that, for example, a cylinder and an annulus would be indistinguishable 
because they are both locally flat and have the same topology. This is, however, not true,
since it is known that in the absence of magnetic flux, the cylinder does 
support Majorana modes, while the annulus does not\cite{read00}. 

The solution to this
puzzle is provided by investigating a cone, where we let the conical angle smoothly
interpolate between the annulus and the cylinder. In this paper, we show that there is a crossover between the ground state and the first-excited state
at a critical opening angle of the cone, and that in general the ground state
supports a current. These effects all depend on the internal geometry of the cone,
but it is also interesting to study effects that depend on the topology. An interesting example is provided by a M\"obius band.

Since a M\"obius band is non-orientable, one would naively think that it cannot
possibly support a chiral superconducting state. On the other hand, for 
a flat, or moderately curved strip, such a state should clearly exist locally, 
which is possible if one allows for line defects between states of different
chirality\cite{Beugeling2014}. Here, we will show that there are two physically different
minimal choices for the line defect, one where the defect is across the M\"obius band, which makes the M\"obius band 
essentially like the disc, and another one for which the defect is along the centre line, which makes the geometry similar to a cylinder. 

The M\"obius band has only one boundary component; hence, the 
response to external fluxes, and the interplay between vortex excitations
and edge Majoranas differ from that of an annulus or a cylinder. We will show that when the defect line is along the centre of the M\"obius band, the Majorana at the defect cannot be removed by the addition of a unit flux. The existence of a Majorana at the line defect is only affected by half a unit of flux passing through the hole of the band. These effects can be understood by mapping the M\"obius band to a corresponding cylinder. Although the existence of the Majorana is not affected by one unit of flux passing through the hole of the band, the profile of the Majorana wave function across the line defect will be modified.

The outline of the paper is the following: in Sec.~\ref{2Dpwave}, we review the physics of a 2d $p$-wave superconductor in the simplest possible setting, namely that of spinless electrons. We begin by introducing the model and then, in the framework of a mean-field approximation, we set the stage for the $p$-wave superconductor. Then, in Sec.~\ref{sec:cone}, we consider a general cone and explain why the non-trivial holonomy of the Levi-Civita connection gives rise to a spontaneous current, and how this affects the Majorana edge modes. Finally, in Secs.~\ref{moebius} and \ref{sec:fluxes} we discuss the M\"obius band and how the Majorana modes in this geometry are affected by inserting flux tubes. Our conclusions are presented in Sec.~\ref{conclusion}.

\section{\texorpdfstring{2$\mathbf d$ p-wave superconductor}{2d p-wave superconductor}} \label{2Dpwave}
Let us start by reviewing the formalism for describing a spinless 2d $p_x+i p_y$ 
superconductor on a general curved surface, and by getting the general
form of the allowed boundary condition at the edges. This is the simplest model for a $p$-wave superconductor in 2d, which allows for a clear presentation of the relevant features.

Since the Pauli principle prevents a local interaction among spinless electrons, the dominating
long wave-length part of any finite-range two-body interaction, which  is only a function
of the geodesic distance\footnote{For any physical system the Hamiltonian would also have terms which does not depend on the intrinsic geometry, but for simplicity we do not consider such terms here.}, can be written as $V(x,y)=\lambda\nabla^{2}\delta^{2}(x-y)$,
where $\nabla^{2}$ is the Laplacian\footnote{For the interaction to be well defined, the Fourier transform
$V(q)$ of the potential $V(r)$, where $r$ is the geodesic distance,
have to vanish as $q\rightarrow\infty$. However, the precise behaviour at large momenta
is not important for the long wave-length physics.}.

The most convenient way to write this interaction in a coordinate-invariant form is to introduce an \emph{orthonormal frame}: a pair of orthonormal unit vectors $\left\{ e_{a}^{\mu}\right\} _{a=1,2}$ at every point. By construction, the metric can be expressed as $g^{\mu\nu} = e_a^\mu e^{a\nu}$, where $e_a^\mu=e^{a\mu}$ by definition.

Performing a partial integration and using that the square of any fermionic operator vanishes, we can express the interaction in second-quantized form as 
\[
\hat{V}=\lambda \int dS\left[\, \left(\psi^\dagger\partial_{-}\psi^\dagger\right)\left(\psi\partial_{+}\psi\right)+\left(\psi^\dagger\partial_{+}\psi^\dagger\right)\left(\psi\partial_{-}\psi\right)\right]\ ,
\]
where $dS=d^{2}x\sqrt{\det g_{\mu\nu}}$, $\partial_{\pm}=e_{1}^{\mu}\partial_{\mu}\pm ie_{2}^{\mu}\partial_{\mu}$ 
and $\psi^\dagger$ and $\psi$ are, respectively, the (spinless) electron creation
and annihilation operators.

In a mean field approach, we approximate the pairing term by
\[
\hat{V}=\frac{\lambda}{2}\int dS\left[   \left(\psi^\dagger\partial_{-}\psi^\dagger\right)\phi_{+}+\left(\psi^\dagger\partial_{+}\psi^\dagger\right)\phi_{-}+h.c.\right] \, ,
\]
where the values of $\phi_{\pm}=\left\langle \psi\partial_{\pm}\psi\right\rangle $ should, strictly speaking, be determined self-consistently. However, we shall take them as given background fields with the correct topological properties. For the 
ground state in the  $p_x + ip_y$ case, we will use $\phi \equiv \phi_+ = const$ and $\phi_- = 0$. 

Since $\phi$ is a vector field with charge $2e$, constant means covariantly constant. 
To this end, we recall that, with respect to the normalized basis $\left\{e_{a}^{\mu}\right\} _{a=1,2}$, the covariant derivative of a general vector with components
$V^{a}$ is
$$D_{\mu}V^{a}=\left(\delta^a_b \partial_{\mu}+\omega_{\mu b}^{\:a}\right)V^{b}\ ,$$
where the connection form is defined by
\begin{equation} 
\omega_{\mu b}^{a}=e_{\nu}^{a}\nabla_{\mu}e^{\nu}_b \label{eq:connection form}\ ,
\end{equation}
with $\nabla_\mu $ the Levi-Civita connection. It can be shown that $\omega_{\mu b}^{a}$ is antisymmetric in $a$ and $b$. 
Therefore, the covariant derivative of  $V_{\pm}:=V^{1}\pm iV^{2}$
is $D_{\mu}V_{\pm}=\left(\partial_{\mu}\pm i\omega_{\mu}\right)V_{\pm}$,
where $\omega_{\mu}\equiv\omega_{\mu2}^{\:1}$. In the presence of an electromagnetic vector potential, this generalises to
\begin{equation} \label{covder}
D_{\mu}\phi=\left(\partial_{\mu}+2ieA_{\mu}+i\omega_{\mu}\right)\phi\ ,
\end{equation} 
where $A_{\mu}$ is the electromagnetic gauge potential. Barring any geometric obstructions, this will vanish in the ground state to minimise the kinetic energy of the system.

The mean-field Hamiltonian reads
\begin{align}
\hat{H}=\int dS  \Psi^{\dagger}\begin{pmatrix}h_{0} & \phi\partial_{-}\\
\phi^{*}\partial_{+} & -h_{0}^{*}
\end{pmatrix}\Psi\ ,
\label{second-quantized-ham}
\end{align}
where $\Psi\equiv\begin{pmatrix}\psi, & \psi^\dagger\end{pmatrix}^{T}$
and the non-interacting part of the Hamiltonian $h_0$ may in general contain any power of the Laplacian. However, since we
are interested only in long-wavelength effects, we truncate it and keep
only the lowest order in derivatives, \emph{i.e.}  we take $h_{0}$ to be
a constant chemical potential,\footnote{
Adding higher-order terms would also mean that the ansatz $\Delta \phi =const.$
is not a self-consistent solution close to vortices or edges.}
 $h_{0}\equiv\mu$. The first-quantized Hamiltonian $\mathcal{H}$
is defined by how it acts on a general single quasi-particle state
\[
|u,v\rangle=\int dS \left(\psi^\dagger,\psi\right)(u,v)^{T}|0\rangle,
\]
where $|0\rangle$ is the ground state, whose existence is assumed. Since the number of quasi-particles is conserved, the Schr\"odinger equation can be rewritten in to a partial differential equation for the functions $u,v$:
\[
i\hbar\frac{\partial}{\partial t}\begin{pmatrix}u\\
v
\end{pmatrix}=\mathcal{H}\begin{pmatrix}u\\
v
\end{pmatrix}\ .
\]
Here, the first quantized Hamiltonian reads 
\begin{equation} 
\mathcal{H}=\begin{pmatrix}-\mu & \frac{1}{2\sqrt{g}}\{\sqrt{g}\phi,\partial_{-}\}\\
-\frac{1}{2\sqrt{g}}\{\sqrt{g}\phi^{*},\partial_{+}\} & \mu
\end{pmatrix}\ ,\label{eq:hamiltonian}
\end{equation}
where $\{ \cdot,\cdot\}$ denotes, as usual, the anti-commutator. The most general local boundary condition for which this Hamiltonian is self-adjoint reads
\begin{equation} \label{bc}
\left.\hat{n}^{\mu}e_{\mu}^{a}\sigma_{a}\begin{pmatrix}\phi^{*}u\\
\phi v
\end{pmatrix}\right|_{\partial S}=\left.s\Delta\begin{pmatrix}u\\
v
\end{pmatrix}\right|_{\partial S},
\end{equation}
where $s$ is an arbitrary real number, $\hat{n}^{\mu}$ is the
outward directed normal, $\Delta=|\phi|$ and $\partial S$ is the edge
of the surface $S$ on which the system is defined. 

The Bogoliubov-deGennes Hamiltonian satisfies
\begin{equation}
	\sigma_{x}\mathcal{H}^{*}\sigma_{x}=\mathcal{H} \ ,\label{eq:symmetry}
\end{equation}
which reflects that in first quantized language, we formally have doubled the degrees of freedom. This property, of the first quantized Hamiltonian, is required for
 $\psi^\dagger$ to be the adjoint of  $\psi$, and is thus not a symmetry, but a consequence of the second-quantized structure. Therefore, only boundary conditions consistent with \eqref{eq:symmetry} are allowed, \emph{i.e.,} $|s|=1$.

\section{Edge Modes on the cone} \label{sec:cone}
\begin{figure}[ht]
	\includegraphics[width=\linewidth]{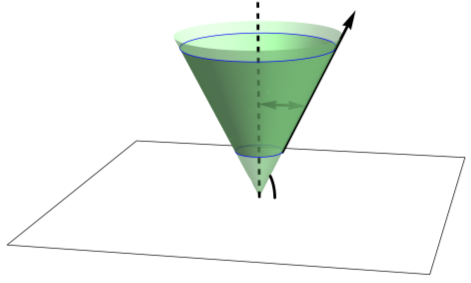}
	\begin{picture}(0,0)
		\put(24,61){$\alpha$}
		\put(28,74){$r=0$}
		\put(38,98){$R(r)$}
		\put(54,128){$r_o$}
		\put(64,145){$r$}
	\end{picture}
	\caption{The embedding given by Eq.~\eqref{embedding}.}
\label{fig:cone}
\end{figure}
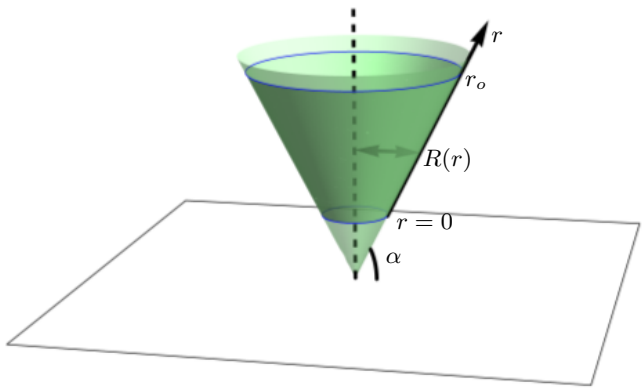
We now specialize to a conical surface, defined by the embedding  
\begin{equation} \label{embedding}
(r,\theta)\mapsto 
\left(R(r)\cos\theta,R(r)\sin\theta,r\sin\alpha\right),
\end{equation}
where $R(r) = R_{i}+r\cos\alpha$, $\theta\in[0,2\pi]$, and $r\in[0,R_o]$ (see Fig.~\ref{fig:cone}). For $\alpha=0$, this
is an annulus with inner  radius $R_{i}$ and outer radius $R_o = R(r_o)$, and for $\alpha=\pi/2$
it is a finite cylinder of radius $R_{i}$ and length $r_o$. The parameter $\alpha$
is the angle that the conic surface makes with the $xy$-plane, and defines 
a smooth interpolation between the annulus and the cylinder.
The metric in the $(r,\theta)$-coordinates is inherited from the Euclidean one by the embedding \eqref{embedding}: 
\begin{equation}
g_{\mu\nu} = \left(\begin{array}{cc}1 & 0 \\0 &   R(r)^2  \end{array}\right) \ .
\end{equation}
In this coordinate system, a natural choice of  orthonormal frame is the set of unit coordinate vectors,
\begin{align} \label{zweibeins}
e_{1} & =\hat{r}\equiv\partial_{r}\ , & e_{2}=\hat{\theta} & \equiv\frac{1} { R(r)}   \partial_{\theta}\ .
\end{align}
These are the normalised radial and angular vectors along the cone. In terms of the Cartesian-coordinate system on the ambient space in which the cone is embedded, we find
\begin {align*}
\hat r &=  (\cos\alpha\cos\theta, \cos\alpha\sin\theta, \sin\alpha) \\
\hat \theta &= (-\sin\theta, \cos\theta,0) \ . 
\end{align*}
We can use the embedding in Eq.~\eqref{embedding} to calculate the connection form, $\omega\equiv\omega^{1}_2$ in Eq.~\eqref{eq:connection form}. It is equal to the Levi-Civita connection for the ambient space projected onto the cone. We obtain
\begin{equation}\label{ConnectionForm}
\omega =d\theta\,  \hat r \cdot \partial_\theta \hat\theta+dr\,  \hat r \cdot \partial_r \hat\theta  = -\cos{\alpha}d\theta \ .\end{equation}
In terms of components this reads ${\omega_\mu=-R(r)\cos(\alpha) e^2_\mu}$, Here $e^2_\mu$ are the components of the 1-form dual to the frame field $e_2$. For a flat surface, the connection is fully determined by the holonomies $\int_
C\omega$ around non-contractible curves. 

For $n$ units of flux through the hole in the cone, and no flux elsewhere, the electromagnetic vector potential reads $ A_\mu=n(2e)^{-1}e^2_\mu$. Due to rotational symmetry, we may take the superconducting order parameter to be
\begin{equation}
\phi = e^{-im\theta} \Delta \, ,
\label{order-param}
\end{equation}
where $\Delta$ is some non-negative real number, and $m$ is an integer. This implies that 
\begin{align}
	\label{cov-dev}
	D_{\mu}\phi = i(n-m-\cos\alpha)e^2_\mu \ .
\end{align}

This expression can only vanish for the annulus  ($\alpha=0$), and for the cylinder ($\alpha=\pi/2$), which means that on a general cone, there is always a non-zero supercurrent present due to the presence of a geometric obstruction.. For the annulus the current vanishes for $m=n-1$, while for the cylinder this happens for $m=n$.

A direct calculation yields
\begin{multline*}
\frac{1}{2\sqrt{g}}\{\sqrt{g}\phi,\partial_{-}\}=\\
e^{-i m\theta}\Delta\left(\partial_{r}+\frac{i}{R(r)}\partial_{\theta}+\frac{m + \cos\alpha}{2R(r)}\right)\ .
\end{multline*}
Substituting the ansatz 
\begin{equation} \label{ansatz}
\begin{pmatrix}u\\
v
\end{pmatrix}=e^{il\theta}\begin{pmatrix}e^{-im\theta/2}u_{l}(r)   \\
e^{im\theta/2}v_{l}(r)
\end{pmatrix}\ 
\end{equation}
into the Hamiltonian Eq.~\eqref{eq:hamiltonian} for integer $l$, and 
eliminating $v_{l}$, we obtain 
\begin{multline}
\left[\mu^{2}-E^{2} - \Delta^2 \left(\partial_{r} - \frac{l- \frac12 \cos\alpha } {R(r) } \right) \right.    \\
\left.  \times \left( \partial_{r}+\frac{l+\frac12 \cos\alpha } {R(r)} \right) \right]  u_{l}(r) = 0 .  \label{eq:diff}
\end{multline}
On the inner edge, the boundary condition \eqref{bc} becomes,
\begin{align*}
v_l(0)&=su_l(0), & s&=\pm 1,
\end{align*}
or by using the equations of motion,
\begin{align}
\left. \left[ s(\mu-E)  + \Delta  \left( \partial_{r}+\frac{l+\frac12 \cos\alpha} {R(r)} \right) \right ]
  u_l(r) \right|_{r=0} \!\!\!\!= 0\, . \label{eq:bc}
\end{align}
For $|E| > \mu$, Eq.~\eqref{eq:diff} has plane-wave solutions, so that $\mu$ is the size of the energy gap. From Eq.~\eqref{eq:diff}, we also see that for $|E| < \mu$ there are two solutions, one which decays with increasing $r$, and one which increases. Which solution is allowed by the boundary condition Eq.~\eqref{eq:bc} depends on the sign of $\mu s$. The decaying solution is only allowed for positive $\mu s$, and the increasing is only allowed for negative $\mu s$. 
Since the solutions have to be normalisable, this reflects that a change of the sign of $\mu$
corresponds to a phase-boundary between a topological and a non-topological phase,
i.e. the phase $\mu s>0$ supports edge modes, while the phase $\mu s<0$ does not\cite{read00}. 

In this analysis, the absolute sign of $\mu$ is of no importance: only the sign relative to the boundary condition, represented by the product $\mu s$,  matters. 
 The boundary condition is, as always, fixed by
the  microscopic physics, \emph{i.e.} by higher-derivative terms. 
Taking $h_{0}=-{\nabla^{2}}/({2M})-\mu$, corresponding to 
a parabolic band, we will, in the 
 $M\rightarrow\infty$  limit, re-obtain the Hamiltonian in Eq.~(\ref{eq:hamiltonian}),
together with the boundary condition $s=1$. Thus, for the usual kinetic term, the topological phase occurs for  $\mu>0$.  

Because of Eq.~(\ref{eq:symmetry}), the spectrum is symmetric around $E=0$. 
For $m=0$, there is a single solution with $l=0$, $E=0$:
\begin{equation} \label{zeromode}
u_{0} =v_0 \sim    \frac 1 {\sqrt{R(r)}}   e^{-s\mu r/\Delta} \ .
\end{equation}
Note that this zero mode is present for any value of the angle $\alpha$, as long as $m\in 2\mathbb Z$.

There is also a zero mode $\sim e^{-\mu(r_o - r)/\Delta}$ located at the other edge. 
Strictly speaking, these solutions are correct only for an infinitely extended cone, \emph{i.e.} $r_o \rightarrow \infty$. However, they
provide a very good approximation, as long as the distance between the edges is large compared to the size $\Delta/\mu$ of the zero mode, since the overlap with the other edge will be exponentially small. 
In the limit $R_i \rightarrow \infty$, the expression for the full spectrum takes the simple form 
\begin{align}
u_l(r) &\sim \exp \left[-\left(\frac{s\mu}\Delta + \frac {\cos\alpha} {2R_i}\right) r  \right],  \\
E_l &= \frac \Delta {R_i}ls\ ,
\end{align}
and again we see that the system contains boundary modes if $\mu s>0$.  Furthermore, the Majorana boundary mode at zero energy occurs only if also $l=0$, which can only happen when $m$ is even.

We now have all the information needed to answer one of the questions 
raised in the introduction -- what happens when we start from 
a cylinder, penetrated by an even number of flux quanta, and slowly squash it into an annulus by  changing the 
angle $\alpha$? The zero mode from Eq.~\eqref{zeromode} remains
as $\alpha$ is decreased towards zero, but the state will now carry current, as we can see from Eq.~\eqref{cov-dev}. By using the order parameter $\phi$ that minimises Eq.~\eqref{cov-dev}, one sees that for $\alpha< \pi/3$, the state with $n-m=1$ carries the smallest current, while for $\alpha> \pi/3$ it is the $n-m=0$ state. Assuming that the energy is a monotonic function of the current, the ground states of the cylinder and the annulus are {\it not} adiabatically connected. At $\alpha=\pi/2$, which coresponds to the cylinder, the zero-current ground state hosts Majorana modes on the edges. Thus, if we make an adiabatic change of $\alpha$, the resulting state on the annulus will have a zero-energy mode, but it will not be the ground state. The ground state, on the other hand will have odd $m$, and thus support no Majorana. However, if we assume that there is a magnetic field, and we allow the system to relax to the ground state by a phase slip in $\phi$, the resulting state on the annulus will enclose an extra flux and the zero-current state will then support a Majorana.


\newcommand{\mob}{M\"obius band}

\section{Edge Modes on the M\"obius band} \label{moebius}

Through the example of a cone, we have shown how inducing a non-local geometric change to a surface can modify the properties of a $p$-wave superconductor. This occurs because the cone has a curvature defect in the hole, which vanishes for the annulus, and reaches $2\pi$ for the cylinder. However, it is not necessary to introduce curvature defects to create surfaces with non-trivial geometric effects. To further explore this possibility, we now turn to the M\"obius band, and use arguments based on the intrinsic geometry to determine the form of the ground state and the number 
and structure of the zero modes at the edge. 

\begin{figure}[ht]
\includegraphics[width=0.5\linewidth]{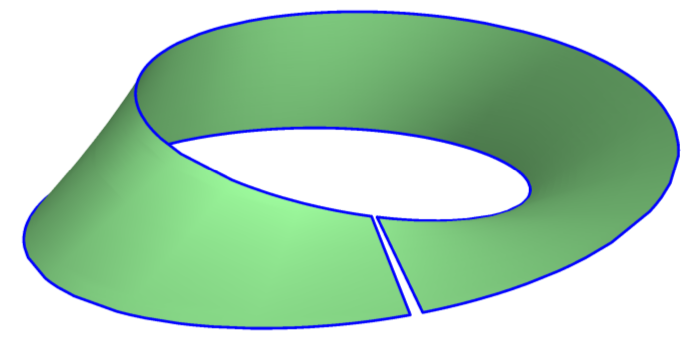}
\caption{\label{mobius2D}  A M\"obius band is formed by gluing together the two edges of a rectangle with a twist. Because one twists the strip before gluing, the M\"obius band is non-orientable and has only a single edge.}
\label{fig:band2}
\end{figure}

Considering only the intrinsic geometry, a flat M\"obius band is a rectangle where the points
of two opposite sides are identified via a reflection. If one considers an embedding in real 3d space, one identifies the two edges after a twist, see Fig. \ref{mobius2D}. 

\begin{figure}[ht]
	\centering
        \subfloat[][]{\includegraphics[width=0.49\linewidth]{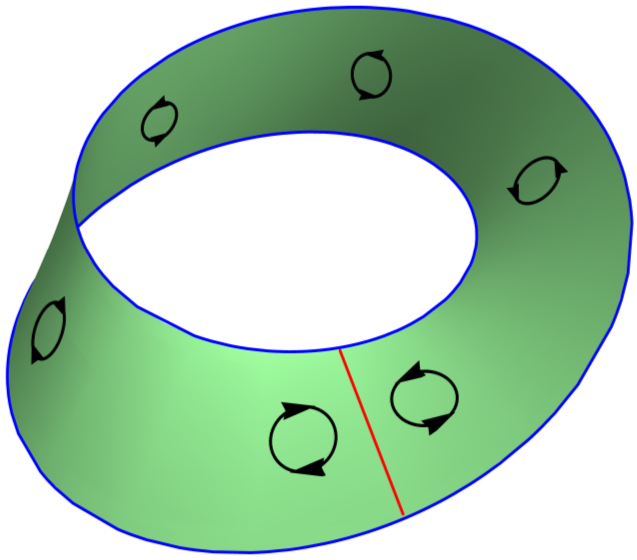}\label{fig:rec-mob}}
	\subfloat[][]{\includegraphics[width=0.49\linewidth]{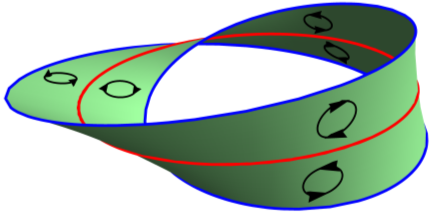}\label{fig:cyl-mob}}
        \caption{Illustration of the two different types of line defects on a M\"obius band supporting a chiral 
        superconductor. The edge of the strip is shown in blue, and the
line defect is shown in red. The circular
arrows indicate the chirality of a $p$-wave superconducting order parameter
that is constant along the strip, except at the defect.}
\label{fig:mob-emb}
\end{figure}

Since the M\"obius band is not orientable, there is no globally defined orthonormal frame $\left\{ e_{a}\right\} _{a=1,2}$. Hence, it is impossible to have a uniform chiral  superconductor on the M\"obius band. However, for large enough systems, 
it is possible to induce such phases {\it locally}, from which
we conclude that the system, by necessity, must exhibit defect lines. 
There are two topologically distinct choices for such a defect line, which 
are depicted in Fig.~\ref{fig:mob-emb}. These are minimal,
 in the sense that you can generate all other domain wall configurations
by either deforming these lines, or by adding new lines surrounding contractible regions.

The configuration that first comes to mind is a defect along
the line in Fig.~\ref{mobius2D}, where the gap is closed by gluing. This results in a M\"obius band with a transverse domain wall, as illustrated in Fig. \ref{fig:rec-mob}. In order to perform a microscopic calculation for this configuration, it is necessary to model the behaviour of the order parameter at the end points of the defect, where two normal edge modes of opposite chirality meet to form a ``double'' mode along the domain wall. We shall not construct such a model here.

It is more interesting to consider a longitudinal line defect, which lies along the centre line of the M\"obius band, as shown in Fig.~\ref{fig:cyl-mob}. By cutting the M\"obius band along the defect line, one obtains a cylindrical geometry (the intrinsic geometry is insensitive to the $4\pi$ twist), as is depicted in Fig.~\ref{fig:flux-hole}. Using this, we can explain the presence of Majorana modes in the M\"obius band by considering the results for the cylinder in the previous section. Imagine that the line defect is not just a line, but is widened to form a thin gap, which we then further widen. Since our model only depends on the intrinsic geometry, there would now be no difference from the canonical cylindrical geometry. In this case, we know when the edge Majorana modes exist from the calculations in section \ref{sec:cone}. Since closing the gap will only affect the physics locally, at one edge of the cylinder, and since the Majorana modes occur and disappear in pairs, the edge Majorana will not disappear when the cylinder is glued back into a M\"obius band. With no flux, and for a large enough strip, we will thus have one zero-energy Majorana on the edge of the M\"obius band and another at the domain wall.

To first order in the derivatives, the edge mode for the M\"obius band will be given by the same solution as for a cylinder, but to better describe the Majorana mode density profile at the line defect, we also need to take into account second-derivative terms in the Hamiltonian. For example, adding a quadratic derivative makes the Majorana wavefunction change under the addition of a superconducting flux-quantum through the hole, even though it does not change the existence of the mode. To be concrete, take
$$
	h_0=-\frac{\nabla^2}{2M}-\mu
$$
in Eq.~\eqref{second-quantized-ham}, where $M$ is the mass of a single Cooper pair. This choice will also explicitly demonstrate the statement made earlier, that higher-order terms select the boundary condition $s=1$ in Eq.~\eqref{bc}. Thus, the requirement for the existence of the Majorana mode is in fact $\mu>0$. 

To construct a coordinate system adapted to the domain wall, we let $x$ denote the length along the domain wall, and $y$ the distance perpendicular to it. If the domain wall has length $L$, the twist in the M\"obius band leads to the boundary condition $(L,y)\sim (0,-y)$. This coordinate system is natural, in the sense that the domain wall is parametrised by the points $(x,0)$. We will take $y\in \mathds{R}$, so that the M\"obius band is infinitely wide, which is equivalent to ignoring the exponentially small splitting that will occur between the Majorana particles due to finite-size effects.

Upon crossing the line defect, the chirality of the superconductor should be flipped, as shown in Fig.~\ref{fig:cyl-mob}. For example, if $\phi_+\neq 0$ on one side of the line defect, then $\phi_-\neq0$ on the other. This is equivalent to changing $\partial_+$ into $\partial_-$ in Eq.~\eqref{second-quantized-ham} and vice versa, when we cross the line defect. This procedure can be effected by replacing $\partial_-$ with 
$$
	\tilde\partial_-:=e_1^{\mu}\partial_\mu-i\operatorname{sgn}(y)e_2^{\mu}\partial_\mu \ ,
$$
and similarly for $\partial_+$. Because the metric is Euclidean, the connection form vanishes and we can make the gauge-choice $\phi=\Delta$. We thus end up with the off-diagonal term (see Eq.~\eqref{eq:hamiltonian})
$$
	\mathcal H_{12}=\frac{1}{2\sqrt{g}}\{\sqrt{g}\phi,\tilde\partial_{-}\}=\Delta\left[ \partial_x +i\operatorname{sgn}(y)\partial_y+i\delta(y)\right]
$$
in the first quantized Hamiltonian $\mathcal H$, which becomes
$$
	\mathcal H=\left(-\frac{\nabla^{2}}{2M}-\mu\right)\sigma^{z}+\Delta\left(\sigma^{x}\partial_{x}+\sigma^{y}\left(\text{sgn}(y)\partial_{y}+\delta(y)\right)\right) \ .
$$

Next, we make the ansatz $\mathcal H (u,v)^T$\-$=0$ to find the solution of the Majorana equation
\begin{align}\label{DomwallAnsatz}
	\begin{pmatrix}e^{-i \pi/4}u(y) \\e^{i \pi/4} v(y)\end{pmatrix}= \left\{ \begin{array}{cl}
	\exp(-\alpha_1 |y|)\chi& \quad y>0\\
	\exp(-\alpha_{-1}|y|)\chi& \quad y<0
	\end{array}\right.\ ,
\end{align}
where $\chi$ is a constant column vector and $\alpha_{\pm1}$ are positive constants. After multiplying the Majorana equation on the left by $\text{diag}(e^{-i\pi/4},e^{i\pi/4})$ and $\sigma_z$, we obtain the two equations (one for $\xi=1$ and one for $\xi=-1$),
\begin{align*}
	0=\left[-\frac{\alpha_\xi^2}{2M}+\frac{\alpha_1+\alpha_{-1}}{2M}\delta(y)-\mu+\Delta(\alpha_\xi -\delta(y))\sigma_y\right]\chi \ .
\end{align*}
Now, put $\chi=\chi_\epsilon$, where $\epsilon=\pm1$ and $\sigma_y\chi_\epsilon=\epsilon\chi_\epsilon$. We then obtain
\begin{subequations}\label{DomEq}
\begin{align}
	0&=-\frac{\alpha_\xi^2}{2M}-\mu+\epsilon\Delta\alpha_\xi\label{DomEq:1}\ ,\\
	\epsilon\Delta&=\frac{\alpha_1+\alpha_{-1}}{2M}\label{DomEq:2} \ .
\end{align}
\end{subequations}
From Eq.~\eqref{DomEq:1}, we find the four possible solutions (recall $\xi=\pm 1$),
\begin{align}
	\alpha_\xi^\pm=-M\left(-\epsilon\Delta\pm\sqrt{\Delta^2-2\frac{\mu}{M}}\right)\ ,
\end{align}
which are positive if and only if $\epsilon=1$ and $\mu>0$. These solutions are two-fold degenerate since they do not depend on $\xi$. Then, Eq.~\eqref{DomEq:2} is satisfied for the two parameter pairs $\{\alpha_1^1,\alpha_{-1}^{-1}\}$ and $\{\alpha_{-1}^1,\alpha_{1}^{-1}\}$ in Eq.~\eqref{DomwallAnsatz}. The solution on the M\"obius band needs to satisfy $u(y)=u(-y)$, since it must obey $u(x+L,y)=u(x,-y)$ because of the boundary condition. Therefore, on the M\"obius band without flux, we have a unique solution, which is the symmetric combination of the two solutions in Eq.~\eqref{DomwallAnsatz},
\begin{align}
	u(y)=-v(y)\propto\left(\exp(-\alpha^1_1 |y|)+\exp(-\alpha^{-1}_1 |y|)\right) \ ,
\end{align}
where we used $\chi_+\propto\left(-e^{i\pi/4},e^{-i\pi/4}\right)^T$.

It is now instructive to take the $M\rightarrow\infty$ limit to obtain the solution without the second order derivative. In this limit, we find 
\begin{align}
	u(y)=-v(y)\propto\exp(-\mu |y|/\Delta)\ ,
\end{align}
which is indeed the solution that you would obtain with boundary condition $s=1$ in Eq.~\eqref{bc}.


\section{Fluxes and vortices in a p-wave M\"obius band}
\label{sec:fluxes}
We already described how the \mob\ in Fig. \ref{fig:band2} can be obtained 
by twisting and gluing a rectangle. From practical experience with 
strips of paper, we expect that such a procedure could be done 
while keeping the geometry locally flat. Since paper does not stretch, any shape one can form out of paper without folding or cutting is intuitively isometric to the plane. Although their expressions are not simple, flat embeddings of the M\"obius band into real 3d space have actually been found.\footnote{
There are many such flat embeddings, and it is interesting to 
ask which one would minimize the elastic energy for a real
physical M{\"o}bius band. This problem was recently solved numerically in Ref.~\onlinecite{StarostinVanDerHeijden2007}.}

In a similar manner, the configuration in Fig.~\ref{fig:cyl-mob}, which has a local chirality, can be formed from an oriented system. We know that cutting a \mob\ along the centre line will produce
a cylinder with a $2\pi$ twist. Thus, by starting from a chiral state on such a 
twisted cylinder, we can fold it into a band, as in Fig.~\ref{fig:cyl-mob}, but with a gap instead of a domain wall. One can then close the gap in a limiting procedure, obtaining a M\"obius band with chirality as depicted in Fig.~\ref{fig:cyl-mob}.

\begin{figure}[htp]
	\centering
        \subfloat[][]{\includegraphics[width=0.49\linewidth]{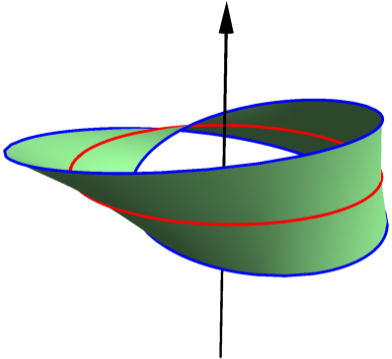}\label{fig:fluxholea}}
	\subfloat[][]{\includegraphics[width=0.49\linewidth]{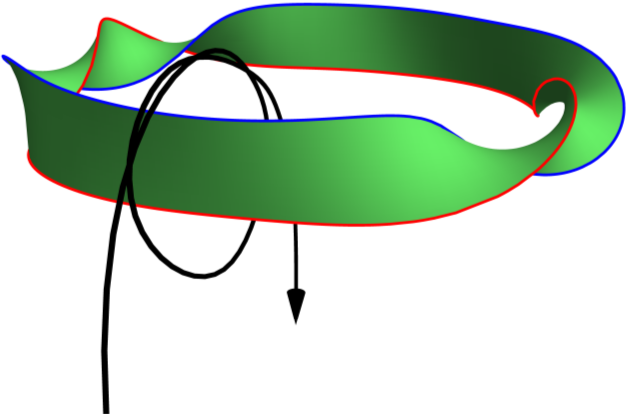}\label{fig:fluxholeb}}
        \caption{(a) The M\"obius band with a domain wall configuration (in red), as in Fig.~\ref{fig:cyl-mob}. The arrow depicts a flux quantum threaded through the hole. (b) The M\"obius band from (a), but cut along the domain wall. After this cut, the band turns into a cylinder with a $4\pi$ twist, and the flux quantum winds around the band twice.}
\label{fig:flux-hole}
\end{figure}

Let us now analyse what happens when a flux is threaded through the
hole of the M\"obius band. We can understand this configuration
by viewing the M\"obius band minus the line defect as an embedding of
a cylinder, as depicted in Fig.~\ref{fig:fluxholeb}. The winding number around the edge of a flux line passing through the hole in the M\"obius strip is two rather than one, and the same holds for the winding number around the line defect. To better understand this statement for the line defect, consider that the latter consists, at each point, of two edge points glued together, and that this doubling also doubles the winding number. Hence, whatever flux we thread through the M\"obius band, it will correspond to twice that amount of flux through the cylinder, as shown in Fig.~\ref{fig:flux-hole}. Therefore, the magnetic flux through the hole of the M\"obius band is not quantized in unit flux quanta, but in half-unit flux quanta\cite{Hayashi2001,Hayashi2005}. Because of this behaviour, the Majoranas on the M\"obius band can be created/destroyed by the addition of half a unit flux through the central hole.  

\begin{figure}[tbp]
	\subfloat[][]{\includegraphics[width=0.49\linewidth]{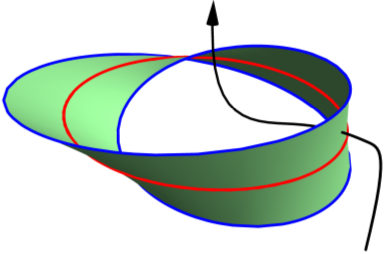}\label{fig:fluxa}}
        \subfloat[][]{\includegraphics[width=0.49\linewidth]{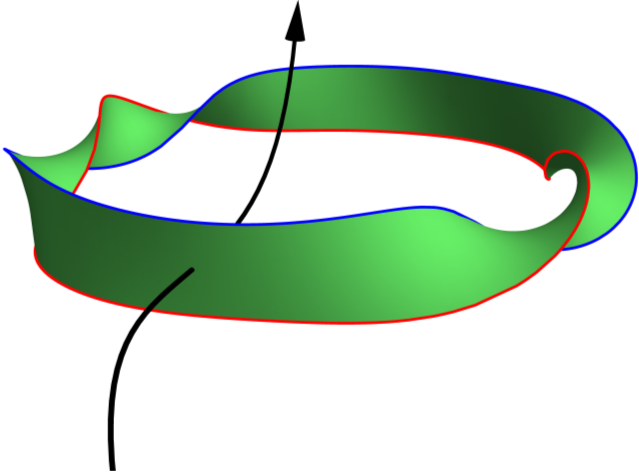}\label{fig:fluxa2}}
	\\
        \subfloat[][]{
                \includegraphics[width=0.49\linewidth]{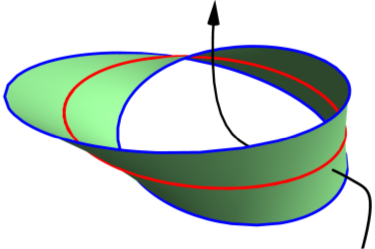}\label{fig:fluxb}}
        \subfloat[][]{\includegraphics[width=0.49\linewidth]{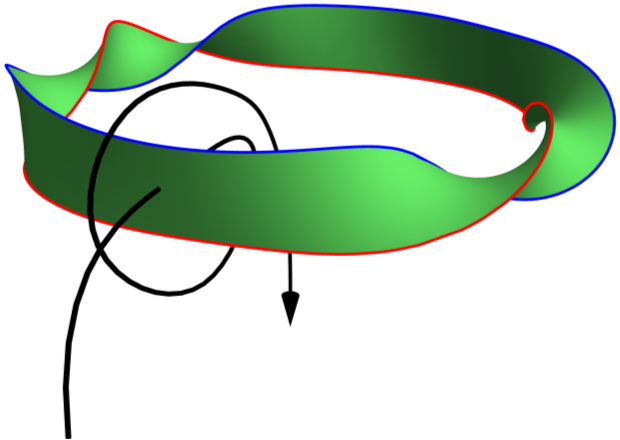}\label{fig:fluxb2}}
        \caption{(a) One of two inequivalent ways to thread a flux through the surface of the M\"obius band. The flux line only encircles the edge, and not the line defect. (b) The situation obtained by cutting the M\"obius band from (a) along the line defect. It is explicitly visible that the flux line winds once around the edge, and not around the line defect. (c) One of two inequivalent ways to thread a flux through the surface of the M\"obius band. The flux line encircles both the edge and the line defect. (d) The situation obtained by cutting the M\"obius band from (c) along the line defect. It is now clear that the flux line winds once around the edge, and twice around the domain wall.}
\label{fig:flux}
\end{figure}

In contrast, one can still only pierce a unit flux through the surface of the band, which will localize a Majorana at the vortex. This must, at the same time, add or remove a Majorana either on the edge or on the line defect. However, as illustrated in Fig. \ref{fig:flux}, no matter how the flux is pierced through, it will always, when the M\"obius band is viewed as a cylinder, wind the edge corresponding to the line defect an even number of times, and thus cannot alter the existence of a Majorana mode there.

As we show now, even though piercing a flux through the M\"obius surface cannot destroy the Majorana at the line defect, adding the quadratic derivative term to the Hamiltonian allows one to distinguish different density profiles for the Majorana mode across the line defect. Adding a flux through the hole of the M\"obius band corresponds to adding a vector potential $A_x=\pi/(eL)$, in the $(x,y)$ coordinate system constructed for the line defect. In the presence of this vector potential, $p_x$ can only vanish for a solution of the form $\exp(-i\pi x/L)\phi(y)$. To fulfil the condition $u(x+L,y)=u(x,-y)$, we need a solution antisymmetric in $y$, and we eventually obtain 
\begin{multline*}
 u(y)=-v(y)\\
\propto e^{-i\pi x/L}\text{sgn}(y)\left(\exp(-\alpha_{1}^{1}|y|)-\exp(-\alpha_{1}^{-1}|y|)\right),
\end{multline*}
which, in the $M\rightarrow\infty$ limit, becomes
\begin{align*}
u(y)=-v(y)\propto\text{sgn}(y)\exp(-\mu |y|/\Delta)\ .
\end{align*}
Notice that for finite $M$ the density profile, \emph{e.g.} $|u(y)|^2$, is different for the symmetric and the anti-symmetric solutions, while for $M\rightarrow\infty$ it is not.

\section{Conclusions}\label{conclusion}

In the ground state, there are Majorana zero-modes present at the edges for a cylindrical geometry, but not for an annulus. We have shown that this apparent inconsistency can be understood by studying how a $p$-wave superconductor is affected by holonomies, \emph{i.e.,} curvature and curvature defects. By tackling this more general problem, we have also described what happens when one continuously deforms a cylinder into an annulus. Deforming the cylinder in this manner will result in an annulus with a supercurrent encircling the hole. This current carrying state will be degenerate due to the presence of Majorana modes, while the ground state will not be.

We have also studied the M\"obius band, which can only host Majorana edge modes in the presence of defect lines. If one chooses a defect line along the centre of the band, it is essentially a cylinder, with the edge corresponding to one side, and the line defect to another. The novel behaviour lies in the response to flux insertion; since the flux lines are mapped in a non-trivial way by the band-cylinder mapping, only specific flux configurations are allowed in the cylinder picture. A flux line through the hole of the band always winds twice in the cylinder picture, and hence these fluxes are quantized at half integers. A flux line through the surface of the band is integer quantized, and winds once around the edge, but twice around the domain wall in the cylinder picture. Consequently, these fluxes do not remove the Majorana at the line defect, but they can change its density profile. Although the results discussed here are mostly of academic interest, and 2d $p$-wave superconductors remain elusive \cite{Maeno2012}, it might be possible to generate $p$-wave superfluids with fermionic ultracold atoms \cite{Cooper2009,Nishida2009}. Moreover, M\"obius bands might also become reality by exploiting a synthetic dimension generated by an internal degree of freedom such as spin in high-spin atoms, and producing a twist via Raman induced hopping at the edges of a ribbon, as suggested recently \cite{Boada2015}. We hope that our results will stimulate experimental efforts n these directions.

\textbf{Acknowledgements:} We thank Giandomenico Palumbo for helpful discussions and for comments on the manuscript. The work by A.Q. and C.M.S. is part of the D-ITP consortium, a program of the Netherlands Organisation for Scientific Research (NWO) that is funded by the Dutch Ministry of Education, Culture and Science (OCW). THH and TK were partially supported by the Swedish Research Council.

\bibliographystyle{apsrev4-1}
\bibliography{p-wave,refs}

\begin{thebibliography}{21}%
\makeatletter
\providecommand \@ifxundefined [1]{%
 \@ifx{#1\undefined}
}%
\providecommand \@ifnum [1]{%
 \ifnum #1\expandafter \@firstoftwo
 \else \expandafter \@secondoftwo
 \fi
}%
\providecommand \@ifx [1]{%
 \ifx #1\expandafter \@firstoftwo
 \else \expandafter \@secondoftwo
 \fi
}%
\providecommand \natexlab [1]{#1}%
\providecommand \enquote  [1]{``#1''}%
\providecommand \bibnamefont  [1]{#1}%
\providecommand \bibfnamefont [1]{#1}%
\providecommand \citenamefont [1]{#1}%
\providecommand \href@noop [0]{\@secondoftwo}%
\providecommand \href [0]{\begingroup \@sanitize@url \@href}%
\providecommand \@href[1]{\@@startlink{#1}\@@href}%
\providecommand \@@href[1]{\endgroup#1\@@endlink}%
\providecommand \@sanitize@url [0]{\catcode `\\12\catcode `\$12\catcode
  `\&12\catcode `\#12\catcode `\^12\catcode `\_12\catcode `\%12\relax}%
\providecommand \@@startlink[1]{}%
\providecommand \@@endlink[0]{}%
\providecommand \url  [0]{\begingroup\@sanitize@url \@url }%
\providecommand \@url [1]{\endgroup\@href {#1}{\urlprefix }}%
\providecommand \urlprefix  [0]{URL }%
\providecommand \Eprint [0]{\href }%
\providecommand \doibase [0]{http://dx.doi.org/}%
\providecommand \selectlanguage [0]{\@gobble}%
\providecommand \bibinfo  [0]{\@secondoftwo}%
\providecommand \bibfield  [0]{\@secondoftwo}%
\providecommand \translation [1]{[#1]}%
\providecommand \BibitemOpen [0]{}%
\providecommand \bibitemStop [0]{}%
\providecommand \bibitemNoStop [0]{.\EOS\space}%
\providecommand \EOS [0]{\spacefactor3000\relax}%
\providecommand \BibitemShut  [1]{\csname bibitem#1\endcsname}%
\let\auto@bib@innerbib\@empty
\bibitem [{\citenamefont {Moore}\ and\ \citenamefont {Read}(1991)}]{moore91}%
  \BibitemOpen
  \bibfield  {author} {\bibinfo {author} {\bibfnamefont {G.}~\bibnamefont
  {Moore}}\ and\ \bibinfo {author} {\bibfnamefont {N.}~\bibnamefont {Read}},\
  }\href {\doibase 10.1016/0550-3213(91)90407-o} {\bibfield  {journal}
  {\bibinfo  {journal} {Nuclear Physics B}\ }\textbf {\bibinfo {volume}
  {360}},\ \bibinfo {pages} {362 } (\bibinfo {year} {1991})}\BibitemShut
  {NoStop}%
\bibitem [{\citenamefont {Read}\ and\ \citenamefont {Rezayi}(1996)}]{read96}%
  \BibitemOpen
  \bibfield  {author} {\bibinfo {author} {\bibfnamefont {N.}~\bibnamefont
  {Read}}\ and\ \bibinfo {author} {\bibfnamefont {E.}~\bibnamefont {Rezayi}},\
  }\href {\doibase 10.1103/PhysRevB.54.16864} {\bibfield  {journal} {\bibinfo
  {journal} {Phys. Rev. B}\ }\textbf {\bibinfo {volume} {54}},\ \bibinfo
  {pages} {16864} (\bibinfo {year} {1996})}\BibitemShut {NoStop}%
\bibitem [{\citenamefont {Read}\ and\ \citenamefont {Rezayi}(1999)}]{read99}%
  \BibitemOpen
  \bibfield  {author} {\bibinfo {author} {\bibfnamefont {N.}~\bibnamefont
  {Read}}\ and\ \bibinfo {author} {\bibfnamefont {E.}~\bibnamefont {Rezayi}},\
  }\href {\doibase 10.1103/PhysRevB.59.8084} {\bibfield  {journal} {\bibinfo
  {journal} {Phys. Rev. B}\ }\textbf {\bibinfo {volume} {59}},\ \bibinfo
  {pages} {8084} (\bibinfo {year} {1999})}\BibitemShut {NoStop}%
\bibitem [{\citenamefont {Read}\ and\ \citenamefont {Green}(2000)}]{read00}%
  \BibitemOpen
  \bibfield  {author} {\bibinfo {author} {\bibfnamefont {N.}~\bibnamefont
  {Read}}\ and\ \bibinfo {author} {\bibfnamefont {D.}~\bibnamefont {Green}},\
  }\href {\doibase 10.1103/PhysRevB.61.10267} {\bibfield  {journal} {\bibinfo
  {journal} {Phys. Rev. B}\ }\textbf {\bibinfo {volume} {61}},\ \bibinfo
  {pages} {10267} (\bibinfo {year} {2000})}\BibitemShut {NoStop}%
\bibitem [{\citenamefont {Greiter}\ \emph {et~al.}(1992)\citenamefont
  {Greiter}, \citenamefont {Wen},\ and\ \citenamefont {Wilczek}}]{greiter92}%
  \BibitemOpen
  \bibfield  {author} {\bibinfo {author} {\bibfnamefont {M.}~\bibnamefont
  {Greiter}}, \bibinfo {author} {\bibfnamefont {X.}~\bibnamefont {Wen}}, \ and\
  \bibinfo {author} {\bibfnamefont {F.}~\bibnamefont {Wilczek}},\ }\href
  {\doibase http://dx.doi.org/10.1016/0550-3213(92)90401-V} {\bibfield
  {journal} {\bibinfo  {journal} {Nuclear Physics B}\ }\textbf {\bibinfo
  {volume} {374}},\ \bibinfo {pages} {567 } (\bibinfo {year}
  {1992})}\BibitemShut {NoStop}%
\bibitem [{\citenamefont {Ivanov}(2001)}]{ivanov01}%
  \BibitemOpen
  \bibfield  {author} {\bibinfo {author} {\bibfnamefont {D.~A.}\ \bibnamefont
  {Ivanov}},\ }\href {\doibase 10.1103/PhysRevLett.86.268} {\bibfield
  {journal} {\bibinfo  {journal} {Phys. Rev. Lett.}\ }\textbf {\bibinfo
  {volume} {86}},\ \bibinfo {pages} {268} (\bibinfo {year} {2001})}\BibitemShut
  {NoStop}%
\bibitem [{\citenamefont {Stern}\ \emph {et~al.}(2004)\citenamefont {Stern},
  \citenamefont {von Oppen},\ and\ \citenamefont {Mariani}}]{stern04}%
  \BibitemOpen
  \bibfield  {author} {\bibinfo {author} {\bibfnamefont {A.}~\bibnamefont
  {Stern}}, \bibinfo {author} {\bibfnamefont {F.}~\bibnamefont {von Oppen}}, \
  and\ \bibinfo {author} {\bibfnamefont {E.}~\bibnamefont {Mariani}},\ }\href
  {\doibase 10.1103/PhysRevB.70.205338} {\bibfield  {journal} {\bibinfo
  {journal} {Phys. Rev. B}\ }\textbf {\bibinfo {volume} {70}},\ \bibinfo
  {pages} {205338} (\bibinfo {year} {2004})}\BibitemShut {NoStop}%
\bibitem [{\citenamefont {{Mourik}}\ \emph {et~al.}(2012)\citenamefont
  {{Mourik}}, \citenamefont {{Zuo}}, \citenamefont {{Frolov}}, \citenamefont
  {{Plissard}}, \citenamefont {{Bakkers}},\ and\ \citenamefont
  {{Kouwenhoven}}}]{kouwenhoven12}%
  \BibitemOpen
  \bibfield  {author} {\bibinfo {author} {\bibfnamefont {V.}~\bibnamefont
  {{Mourik}}}, \bibinfo {author} {\bibfnamefont {K.}~\bibnamefont {{Zuo}}},
  \bibinfo {author} {\bibfnamefont {S.~M.}\ \bibnamefont {{Frolov}}}, \bibinfo
  {author} {\bibfnamefont {S.~R.}\ \bibnamefont {{Plissard}}}, \bibinfo
  {author} {\bibfnamefont {E.~P.~A.~M.}\ \bibnamefont {{Bakkers}}}, \ and\
  \bibinfo {author} {\bibfnamefont {L.~P.}\ \bibnamefont {{Kouwenhoven}}},\
  }\href {\doibase 10.1126/science.1222360} {\bibfield  {journal} {\bibinfo
  {journal} {Science}\ }\textbf {\bibinfo {volume} {336}},\ \bibinfo {pages}
  {1003} (\bibinfo {year} {2012})},\ \Eprint {http://arxiv.org/abs/1204.2792}
  {arXiv:1204.2792 [cond-mat.mes-hall]} \BibitemShut {NoStop}%
\bibitem [{\citenamefont {Churchill}\ \emph {et~al.}(2013)\citenamefont
  {Churchill}, \citenamefont {Fatemi}, \citenamefont {Grove-Rasmussen},
  \citenamefont {Deng}, \citenamefont {Caroff}, \citenamefont {Xu},\ and\
  \citenamefont {Marcus}}]{marcus13}%
  \BibitemOpen
  \bibfield  {author} {\bibinfo {author} {\bibfnamefont {H.~O.~H.}\
  \bibnamefont {Churchill}}, \bibinfo {author} {\bibfnamefont {V.}~\bibnamefont
  {Fatemi}}, \bibinfo {author} {\bibfnamefont {K.}~\bibnamefont
  {Grove-Rasmussen}}, \bibinfo {author} {\bibfnamefont {M.~T.}\ \bibnamefont
  {Deng}}, \bibinfo {author} {\bibfnamefont {P.}~\bibnamefont {Caroff}},
  \bibinfo {author} {\bibfnamefont {H.~Q.}\ \bibnamefont {Xu}}, \ and\ \bibinfo
  {author} {\bibfnamefont {C.~M.}\ \bibnamefont {Marcus}},\ }\href {\doibase
  10.1103/PhysRevB.87.241401} {\bibfield  {journal} {\bibinfo  {journal} {Phys.
  Rev. B}\ }\textbf {\bibinfo {volume} {87}},\ \bibinfo {pages} {241401}
  (\bibinfo {year} {2013})}\BibitemShut {NoStop}%
\bibitem [{\citenamefont {{Beugeling}}\ \emph {et~al.}(2014)\citenamefont
  {{Beugeling}}, \citenamefont {{Quelle}},\ and\ \citenamefont {{Morais
  Smith}}}]{Beugeling2014}%
  \BibitemOpen
  \bibfield  {author} {\bibinfo {author} {\bibfnamefont {W.}~\bibnamefont
  {{Beugeling}}}, \bibinfo {author} {\bibfnamefont {A.}~\bibnamefont
  {{Quelle}}}, \ and\ \bibinfo {author} {\bibfnamefont {C.}~\bibnamefont
  {{Morais Smith}}},\ }\href {\doibase 10.1103/PhysRevB.89.235112} {\bibfield
  {journal} {\bibinfo  {journal} {Phys. Rev. B}\ }\textbf {\bibinfo {volume}
  {89}},\ \bibinfo {eid} {235112} (\bibinfo {year} {2014})}\BibitemShut
  {NoStop}%
\bibitem [{Note1()}]{Note1}%
  \BibitemOpen
  \bibinfo {note} {For any physical system the Hamiltonian would also have
  terms which does not depend on the intrinsic geometry, but for simplicity we
  do not consider such terms here.}\BibitemShut {Stop}%
\bibitem [{Note2()}]{Note2}%
  \BibitemOpen
  \bibinfo {note} {For the interaction to be well defined, the Fourier
  transform $V(q)$ of the potential $V(r)$, where $r$ is the geodesic distance,
  have to vanish as $q\rightarrow \infty $. However, the precise behaviour at
  large momenta is not important for the long wave-length physics.}\BibitemShut
  {Stop}%
\bibitem [{Note3()}]{Note3}%
  \BibitemOpen
  \bibinfo {note} {Adding higher-order terms would also mean that the ansatz
  $\Delta \phi =const.$ is not a self-consistent solution close to vortices or
  edges.}\BibitemShut {Stop}%
\bibitem [{Note4()}]{Note4}%
  \BibitemOpen
  \bibinfo {note} {There are many such flat embeddings, and it is interesting
  to ask which one would minimize the elastic energy for a real physical
  M{\"o}bius band. This problem was recently solved numerically in
  Ref.~\protect \rev@citealpnum {StarostinVanDerHeijden2007}.}\BibitemShut
  {Stop}%
\bibitem [{\citenamefont {Hayashi}\ and\ \citenamefont
  {Ebisawa}(2001)}]{Hayashi2001}%
  \BibitemOpen
  \bibfield  {author} {\bibinfo {author} {\bibfnamefont {M.}~\bibnamefont
  {Hayashi}}\ and\ \bibinfo {author} {\bibfnamefont {H.}~\bibnamefont
  {Ebisawa}},\ }\href {\doibase 10.1143/JPSJ.70.3495} {\bibfield  {journal}
  {\bibinfo  {journal} {Journal of the Physical Society of Japan}\ }\textbf
  {\bibinfo {volume} {70}},\ \bibinfo {pages} {3495} (\bibinfo {year}
  {2001})}\BibitemShut {NoStop}%
\bibitem [{\citenamefont {Hayashi}\ \emph {et~al.}(2005)\citenamefont
  {Hayashi}, \citenamefont {Ebisawa},\ and\ \citenamefont
  {Kuboki}}]{Hayashi2005}%
  \BibitemOpen
  \bibfield  {author} {\bibinfo {author} {\bibfnamefont {M.}~\bibnamefont
  {Hayashi}}, \bibinfo {author} {\bibfnamefont {H.}~\bibnamefont {Ebisawa}}, \
  and\ \bibinfo {author} {\bibfnamefont {K.}~\bibnamefont {Kuboki}},\ }\href
  {\doibase 10.1103/PhysRevB.72.024505} {\bibfield  {journal} {\bibinfo
  {journal} {Phys. Rev. B}\ }\textbf {\bibinfo {volume} {72}},\ \bibinfo
  {pages} {024505} (\bibinfo {year} {2005})}\BibitemShut {NoStop}%
\bibitem [{\citenamefont {Maeno}\ \emph {et~al.}(2012)\citenamefont {Maeno},
  \citenamefont {Kittaka}, \citenamefont {Nomura}, \citenamefont {Yonezawa},\
  and\ \citenamefont {Ishida}}]{Maeno2012}%
  \BibitemOpen
  \bibfield  {author} {\bibinfo {author} {\bibfnamefont {Y.}~\bibnamefont
  {Maeno}}, \bibinfo {author} {\bibfnamefont {S.}~\bibnamefont {Kittaka}},
  \bibinfo {author} {\bibfnamefont {T.}~\bibnamefont {Nomura}}, \bibinfo
  {author} {\bibfnamefont {S.}~\bibnamefont {Yonezawa}}, \ and\ \bibinfo
  {author} {\bibfnamefont {K.}~\bibnamefont {Ishida}},\ }\href {\doibase
  10.1143/JPSJ.81.011009} {\bibfield  {journal} {\bibinfo  {journal} {Journal
  of the Physical Society of Japan}\ }\textbf {\bibinfo {volume} {81}},\
  \bibinfo {pages} {011009} (\bibinfo {year} {2012})},\ \Eprint
  {http://arxiv.org/abs/http://dx.doi.org/10.1143/JPSJ.81.011009}
  {http://dx.doi.org/10.1143/JPSJ.81.011009} \BibitemShut {NoStop}%
\bibitem [{\citenamefont {Cooper}\ and\ \citenamefont
  {Shlyapnikov}(2009)}]{Cooper2009}%
  \BibitemOpen
  \bibfield  {author} {\bibinfo {author} {\bibfnamefont {N.~R.}\ \bibnamefont
  {Cooper}}\ and\ \bibinfo {author} {\bibfnamefont {G.~V.}\ \bibnamefont
  {Shlyapnikov}},\ }\href {\doibase 10.1103/PhysRevLett.103.155302} {\bibfield
  {journal} {\bibinfo  {journal} {Phys. Rev. Lett.}\ }\textbf {\bibinfo
  {volume} {103}},\ \bibinfo {pages} {155302} (\bibinfo {year}
  {2009})}\BibitemShut {NoStop}%
\bibitem [{\citenamefont {{Nishida}}(2009)}]{Nishida2009}%
  \BibitemOpen
  \bibfield  {author} {\bibinfo {author} {\bibfnamefont {Y.}~\bibnamefont
  {{Nishida}}},\ }\href {\doibase 10.1016/j.aop.2008.10.011} {\bibfield
  {journal} {\bibinfo  {journal} {Annals of Physics}\ }\textbf {\bibinfo
  {volume} {324}},\ \bibinfo {pages} {897} (\bibinfo {year}
  {2009})}\BibitemShut {NoStop}%
\bibitem [{\citenamefont {{Boada}}\ \emph {et~al.}(2015)\citenamefont
  {{Boada}}, \citenamefont {{Celi}}, \citenamefont {{Rodr{\'{\i}}guez-Laguna}},
  \citenamefont {{Latorre}},\ and\ \citenamefont {{Lewenstein}}}]{Boada2015}%
  \BibitemOpen
  \bibfield  {author} {\bibinfo {author} {\bibfnamefont {O.}~\bibnamefont
  {{Boada}}}, \bibinfo {author} {\bibfnamefont {A.}~\bibnamefont {{Celi}}},
  \bibinfo {author} {\bibfnamefont {J.}~\bibnamefont
  {{Rodr{\'{\i}}guez-Laguna}}}, \bibinfo {author} {\bibfnamefont {J.~I.}\
  \bibnamefont {{Latorre}}}, \ and\ \bibinfo {author} {\bibfnamefont
  {M.}~\bibnamefont {{Lewenstein}}},\ }\href {\doibase
  10.1088/1367-2630/17/4/045007} {\bibfield  {journal} {\bibinfo  {journal}
  {New Journal of Physics}\ }\textbf {\bibinfo {volume} {17}},\ \bibinfo {eid}
  {045007} (\bibinfo {year} {2015})}\BibitemShut {NoStop}%
\bibitem [{\citenamefont {Starostin}\ and\ \citenamefont {van~der
  Heijden}(2007)}]{StarostinVanDerHeijden2007}%
  \BibitemOpen
  \bibfield  {author} {\bibinfo {author} {\bibfnamefont {E.~L.}\ \bibnamefont
  {Starostin}}\ and\ \bibinfo {author} {\bibfnamefont {G.~H.~M.}\ \bibnamefont
  {van~der Heijden}},\ }\href {\doibase 10.1038/nmat1929} {\bibfield  {journal}
  {\bibinfo  {journal} {Nature Mater.}\ }\textbf {\bibinfo {volume} {6}},\
  \bibinfo {pages} {563} (\bibinfo {year} {2007})}\BibitemShut {NoStop}%
\end{thebibliography}%

\end{document}